\documentclass[
reprint,
nofootinbib,
 amsmath,amssymb,
 aps,
]{revtex4-2}
\usepackage[dvipsnames,table,xcdraw]{xcolor}
\usepackage{graphicx}
\usepackage{bm}
\usepackage{tikz}
\usepackage{braket}
\usepackage{dsfont}
\usepackage{hyperref}
\hypersetup{
    colorlinks=true, 
    linktoc=all,     
    linkcolor=blue,  
    citecolor=magenta
}
\usepackage{bbold}
\usepackage{soul}

\usetikzlibrary{decorations.pathreplacing,calligraphy}
\usetikzlibrary{decorations.markings}
\newcommand{\mysection}[1]{\paragraph*{#1 ---}}

\newcommand{\hone}{H_1}

{\renewcommand{\arraystretch}{1.2}}

\makeatletter

\begin{document}

\preprint{APS/123-QED}

\title{Entanglement Oscillations from Many-Body Quantum Scars}

\author{Nicholas O'Dea}
\thanks{Nicholas O'Dea and Adithya Sriram contributed equally to this work.}
\author{Adithya Sriram}
\thanks{Nicholas O'Dea and Adithya Sriram contributed equally to this work.}
\affiliation{Department of Physics, Stanford University, Stanford, CA 94305, USA}

\begin{abstract}
Quantum scars are nonthermal eigenstates that prevent thermalization of initial states with weight on the scars. When the scar states are equally spaced in energy, superpositions of scars show oscillating local observables that can be detected in experiments. However, we note that scarred models in the literature show fundamentally different scar entanglement dynamics: some show entanglement oscillations while others are completely frozen. We explain this freezing through a no-go theorem which we apply to more than a dozen scarred models in the literature. We also discuss a method for evading the no-go theorem by deforming the scarred models with frozen scar entanglement dynamics.
\end{abstract}

\maketitle

\mysection{Introduction} The advent of quantum simulators and new experimental platforms has motivated the study of different types of many-body quantum phenomena. These platforms have also uncovered surprises, such as models with non-thermal behavior~\cite{bernien2017probing} like periodic revivals of observables in special initial states otherwise expected to rapidly thermalize. The state-dependent nonthermal dynamics in this Rydberg atom experiment were identified~\cite{turner2018scars} with the existence of approximate ``scar" states with outlier local expectation values in the simulated PXP model, prompting a flurry of theoretical investigation into this thermalization-evading phenomenon (see~\cite{Serbyn-Papic2021_review,Moudgalya-Regnault2021_review,Chandran-Moessner2022_review} for literature reviews) and experimental realizations of PXP and other scarred models~\cite{kao2021pumping,bluvstein2021controlling,jepsen2022heisenberg,chen2022mitigated,su2023bose,zhang2023many,yang2024phantom}. Because of their outlier local expectation values, scar states violate the eigenstate thermalization hypothesis~\cite{Jensen:1985aa, Deutsch1, Sred1, Rigol:2008bh, DAlessio:2016aa, nandkishorehuse2015_mbl_review, abanin2019_mbl_review} in a weak sense and hence allow initial states with weight on the scar states to evade thermalization. When the scar states are equally split in energy (i.e. form a ``scar tower"), superpositions of scars will show periodic motion that can be probed through oscillations in observables. 

The special initial state in the PXP model also shows non-monotonic entanglement dynamics~\cite{turner2018scars,wwho2019_periodicorbits}, probed experimentally in another Rydberg atom experiment~\cite{bluvstein2022processor}. Though analysis of the PXP model is complicated by the fact that its scar states are inexact, deformations of the PXP model that improve the lifetimes of the approximate scars show persistent, periodic oscillations in entanglement \cite{choi2019perfect, abanin2020slow, ljubotina2022optimal}. Investigations into many models of exact scars were motivated because of the PXP model. However, we argue that oscillations in entanglement entropy starting from quenches to superpositions of scars are in fact forbidden in many models of exact scars -- any superposition of scars in these models has all measures of entanglement frozen in time. This frozen entanglement is not a defect of these models relative to models that show scar entanglement dynamics, but instead demonstrates differences in the phenomenology of scarred models.

Namely, we prove that scar entanglement dynamics cannot occur whenever the scars' energies can be replicated by an operator $\hone$ that is a sum of single-site operators. This occurs quite commonly; see the first 14 entries of Table \ref{tab:scartable}. We also show how \textit{deformations} of these models by unitary rotations can induce entanglement oscillations. Our results help elucidate one aspect of the model and experiments which unleashed this flurry of research into quantum many body scars.

\mysection{No-go theorem for entanglement dynamics} Consider a set of states $\{ \ket{\phi_n} \}$ and a Hamiltonian $H$ that has these states as eigenstates; $H \ket{\phi_n} = E_n \ket{\phi_n}$. Suppose there exists a sum of single-site operators $\hone = \sum_j h_j$ that also reproduces these energies, $\hone\ket{\phi_n} = E_n \ket{\phi_n}$. Then all superpositions of these states $\sum_n c_n \ket{\phi_n}$ will have time-independent entanglement under the dynamics generated by $H$.\footnote{The value of the entanglement may depend on the $c_n$, but it will be independent of time under the dynamics generated by $H$.}

To see this, note that a state $\ket{\psi} = \sum_n c_n \ket{\phi_n}$ will evolve under $H$ as
\begin{equation}
\begin{split} \label{eq:eigenstate_decomp}
    \ket{\psi(t)} = e^{-iHt} \ket{\psi} &= \sum_n c_n e^{-i E_n t} \ket{\phi_n}  \\ &= e^{-i\hone t} \ket{\psi}.
\end{split}
\end{equation}
However, by the definition of $\hone$, $e^{-i\hone t} = \otimes_j e^{-i h_j t}$; that is, $e^{-iH_1 t}$ decomposes into a product of single-site unitaries and cannot change the entanglement of any state. Since $\ket{\psi(t)} = e^{-i\hone t} \ket{\psi}$, the entanglement of $\ket{\psi(t)}$ is time-independent. In particular, all entanglement monotones of $\ket{\psi(t)}$ cannot change with time, including the von Neumann and Renyi entanglement entropies across any cut.

It also suffices for $H_1$ to reproduce the energies up to a constant energy shift $\tilde{E}$ (i.e. $\hone\ket{\phi_n} = (E_n+\tilde{E}) \ket{\phi_n}$), as a global phase cannot change entanglement. We will freely omit such constant shifts when exhibiting $H_1$ for various models in the following.

While the theorem above is perfectly general, we will specialize to scar states $|\phi_n\rangle$ and scarred Hamiltonians $H$ in the following. In Table \ref{tab:scartable}, we collect more than a dozen families of scarred models that lack entanglement dynamics because of the existence of an $H_1$ satisfying the conditions of the no-go theorem.

\mysection{Examples}
The spin-$1$ AKLT model is an $SO(3)$-symmetric Hamiltonian
\begin{equation}
    H_{\rm AKLT} = \sum_{i=1}^L P^{(2)}_{i,i+1}
\end{equation}
that is a sum of projectors onto two-site spin-$2$ configurations. For ease, we will restrict to an even length $L$. In periodic boundary conditions, it has a unique spin-$0$ ground state $\ket{g}$; the states $|\phi_n\rangle \propto (Q^\dagger)^n \ket{g}$ with $Q^\dagger = \sum_{j=1}^L (-1)^j (S_j^+)^2$ form an equally spaced scar tower with energies $2n$. A particular superposition of these scar states, $|\psi\rangle \propto e^{Q^\dagger}\ket{g}$, was proposed~\cite{marklinmotrunich2020unified} as a good initial condition for experiments to see oscillations in observables.

The states in this tower are all eigenstates of $H_1 \sum_{i=1}^L S^z_i$, which reproduces the energy of the scars. By our no-go theorem, this means that any and every superposition of the states $(Q^\dagger)^n \ket{g}$ (including $e^{Q^\dagger}\ket{g}$) will not show entanglement dynamics under $H_{\rm AKLT}$. Although local observables (like $(S^+_i)^2 + h.c.$) will show oscillations, the entanglement will show no dynamics at all. In Appendix~\ref{app:AKLT}, we consider a more involved case that uses all the scars of the model, including those related by $SO(3)$ symmetry; i.e. superpositions of states of the form $(S^-)^m (Q^\dagger)^n \ket{g}$. 

As another example, consider the spin-$1/2$ ``domain wall conserving model" \cite{iadecola2020kinetic}:
\begin{align}\label{eq:dwc}
    H = \sum_{i = 1}^L \left[ \lambda (\sigma_i^x - \sigma_{i-1}^z \sigma_i^x \sigma_{i+1}^z) + \Delta \sigma_i^z + J\sigma_i^z \sigma_{i+1}^z \right]
\end{align}
This model has two different raising operators~\cite{marklinmotrunich2020unified} $R^\dagger$ and $P^\dagger$ giving rise to a ``pyramid" of scars $\ket{\phi_{n,m}} \propto (P^\dagger)^m (R^\dagger)^n |\downarrow ... \downarrow\rangle$ with energies $(2 \Delta - 4J)n + 2\Delta m$ up to a constant shift. The explicit forms of $R^\dagger$ and $P^\dagger$ are not central to our purposes. Both increase $\sum_i \sigma^z_i$ by $2$, while only $R^\dagger$ changes the number of domain walls (by $2$). In particular, $\sum_{i=1}^L \sigma_i^z \ket{\phi_{n,m}} = (2n+2m-L)\ket{\phi_{n,m}}$. 

The state $e^{c R^\dagger}\ket{\phi_{0,0}}$ was proposed in~\cite{iadecola2020kinetic} as a possible initial state. However, it is a superposition of the states $\ket{\phi_{n,0}}$, and the choice of $H_1 = \Delta \sum_{i=1}^L \sigma_i^z$ reproduces the energies of these states. This forbids entanglement dynamics under $H$ starting in any superposition of these states, including $e^{c R^\dagger}\ket{\phi_{0,0}}$. We can find choices of $H_1$ satisfying the conditions of the no-go theorem more generally:  ``cuts" along the pyramid consisting of $\ket{\phi_{n,m}}$ with $m$ fixed have $H_1 = \Delta \sum_{i=1}^L \sigma_i^z$, while cuts consisting of $\ket{\phi_{n,m}}$ with $n$ fixed have $H_1 = (\Delta - 2 J) \sum_{i=1}^L \sigma_i^z$. 

\begin{figure}[h!]
    \centering
    \includegraphics[width=\linewidth]{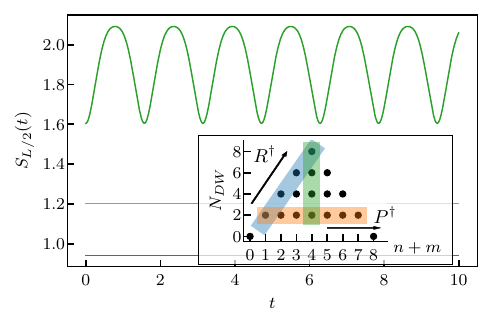}
    \caption{Half-chain entanglement dynamics from scar pyramid states in the DWC model in Eq.~\ref{eq:dwc}. The half-chain entanglement $S_{L/2}(t)$ is plotted for initial states that are uniform superpositions of states within three slices of the scar pyramid. Only the vertical slice which has constant $\sum \sigma^z_i$ but is split in energy by $\sum \sigma^z_i \sigma^z_{i+1}$ exhibits oscillations. The inset shows the scar pyramid (based on that in~\cite{marklinmotrunich2020unified}) and the respective states involved in the superpositions that we time evolve. The $y$-axis of the inset is the number $N_{DW}$ of domain walls in the state, and the $x$ axis is $n+m$ (see discussion surrounding Eq.~\ref{eq:dwc}). We use $L = 8$ and Hamiltonian parameters $\lambda = 1, \Delta = 0.1$ and $J = 1$. The results are analogous if random superpositions of scars rather than uniform superpositions are used for the initial state. }
    \label{fig:pyramid_oscillations}
\end{figure}

However, consider a different cut of scar states with fixed $n+m$ (with $0<n+m<L$); such states are within a fixed $\sum_i \sigma_i^z$ symmetry sector. $\sum_i \sigma_i^z$ is the only candidate operator for $H_1$, as it is the only sum of single-site operators that has such scar states as eigenstates. However, for fixed $n+m$, $\sum_i \sigma_i^z$ is also fixed and cannot reproduce the eigenvalues of $H$. Superpositions of such states within the same $\sum_{i=1}^L \sigma_i^z$ sector thus evade the no-go theorem and indeed show entanglement oscillations in Fig.~\ref{fig:pyramid_oscillations}.

\bgroup
\squeezetable
\begin{table}[]
    \centering
    \begin{ruledtabular}
    \def\arraystretch{1.3}
    \begin{tabular}{ p{3 cm} | p{5 cm} }
    \textbf{Model} & \textbf{Entanglement Oscillations} \\ \hline
    Spin-1 XY~\cite{iadecola2019xy, marklinmotrunich2020unified}, 
 $q$-deformed XY~\cite{odea2020tunnels} & \textcolor{red}{No,} by $\hone \propto \sum_i S_i^z$.
 \\ \hline
    Bond-bimagnon~\cite{chattopadhyay2020bondbimagnon, iadecola2019xy}  & \textcolor{red}{No,} by $\hone \propto \sum_i S_i^z$. \\ \hline
     Spin-1/2 maximal Casimir \cite{choi2019perfect}, including DMI model~\cite{markmotrunich2020}   & \textcolor{red}{No,} by $\hone \propto \sum_i S_i^z$. \\ \hline
    Non-maximal Casimir \cite{odea2020tunnels} & \textcolor{red}{No,} by $\hone \propto \sum_i S_i^z$. \\ \hline
    Onsager algebra clock model \cite{PhysRevLett.124.180604} & \textcolor{red}{No,} by $\hone \propto \sum_i S_i^z$.  \\ \hline
    Kagome XXZ \cite{lee2020threecolor} &    \textcolor{red}{No,} by $\hone \propto \sum_i S_i^z$.
    \\ \hline Motif Hamiltonians~\cite{chertkov2021motif} &  \textcolor{red}{No,} by $\hone \propto \sum_i S_i^z$. \\ \hline
    Perturbed Potts \cite{moudgalya2020mps} & \textcolor{red}{No}, by $\hone \propto \sum_i S_i^z$. 
    \\ \hline
    Rainbow tower~\cite{langlett2022rainbow} & \textcolor{red}{No}, by $\hone \propto \sum_i S_i^z$.
    \\ \hline
    Topological tower~\cite{ren2022deformed} & \textcolor{red}{No}, by $\hone \propto \sum_i S_i^z$. \\ \hline
    Multi-$\pi$-magnon~\cite{tang2022multimagnon} & \textcolor{red}{No}, by $\hone \propto \sum_i S_i^z$. \\ \hline
    Infinite EPR tower ~\cite{wildeboer2022EPR} & \textcolor{red}{No}, by $\hone \propto \sum_i (n_i \otimes I + I \otimes n_i)$. \\ \hline
    Maximal $SU(3)$ Casimir \cite{odea2020tunnels, ren2021quasisymmetry} & \textcolor{red}{No,} by $\hone = \sum_i a S^z_i + b (S^z_i)^2$ for appropriate constants $a$ and $b$. \\ \hline
    Generalized Hubbard Models \cite{10.21468/SciPostPhys.3.6.043, markmotrunich2020, PhysRevB.102.085140, PhysRevLett.63.2144, pakrouski2020invariant} & \textcolor{red}{No,} by $H_1 \propto \sum_{j,\sigma} n_{j,\sigma}$. \\ \hline

Domain Wall Conserving Model \cite{marklinmotrunich2020unified, iadecola2020kinetic} & \textcolor{ForestGreen}{Yes}, but requires using both ladder operators (see \textit{Examples}), else \textcolor{red}{no} by $\hone \propto \sum_i S_i^z$. \\ \hline

Spin-1 AKLT \cite{moudgalya2018nonintegrable, moudgalya2018entanglement, marklinmotrunich2020unified, moudgalya2020mps} &  \textcolor{ForestGreen}{Yes}, but requires using both ladder operators (see \textit{Examples} and Appendix~\ref{app:AKLT}), else \textcolor{red}{no} by $\hone \propto \sum_i S_i^z$. Some entanglement cuts are frozen. \\ \hline

Correlated-hopping Bose-Hubbard ($H_3$ in \cite{hudomal2020correlated}) &  \textcolor{ForestGreen}{Yes}, but some entanglement cuts are frozen. \\ \hline
    \end{tabular}
    \end{ruledtabular}

    \caption{Non-exhaustive table of models with exact, equally spaced scars taken from the literature. The domain-wall conserving model, AKLT, and maximal $SU(3)$ Casimir models have ``pyramids" of scars generated by two scar creation operators. We list the $H_1$ that by the no-go theorem will prevent entanglement oscillations in any superpositions of the scars, whenever such an operator exists. }
    \label{tab:scartable}
\end{table}
    
\egroup

\mysection{Generalization} There is a useful generalization of the no-go theorem with both weaker assumptions and weaker consequences. For ease, we will first restrict to the special case of spin chains. Suppose there is an $H_k$ which is a sum of commuting $k$-body terms (with the $k$ sites contiguous) such that $H$ and $H_k$ share the same eigenvalues on the scar subspace. Suppose the entanglement measure of interest is the entanglement entropy between contiguous regions $A$ and $B$. Then time-evolving a superposition of scars under $H$ will have a bound on entropy oscillations of the form $\max_t S(t) - \min_t S(t) \leq 2(k-1) \log(2)$. This is because only those terms of $H_k$ that straddle the cut between the regions will be able to contribute to changes in the entanglement entropy. For the domain wall conserving model, $H_2 = \sum_i \Delta \sigma_i^z + J \sigma_i^z \sigma_{i+1}^z$ satisfies the conditions above, and hence the amplitude of entanglement oscillations in time is at most $2 \log(2)$ starting in any superposition of the $\ket{\phi_{n,m}}$. In higher dimensions $d>1$, more terms in $H_k$ can straddle the boundary between $A$ and $B$, giving area-law oscillations that are bounded by $O((k-1)^d)$ times the size of the boundary between $A$ and $B$.

\mysection{Scars from Cartan subalgebra of onsite symmetries}
The literature on constructing scar towers from symmetries often \textit{explicitly uses} such operators $H_1$ to generate $H$'s with equal energy splittings between scars, directly making such $H$'s unable to show scar entanglement dynamics. 

For example, one of the constructions in \cite{odea2020tunnels} uses the Cartan subalgebra of an explicitly broken non-abelian symmetry to split the scars in energy; such terms yield the spectrum-generating part $H_{SG}$ of the scarred Hamiltonian $H$ ($H|\phi_n\rangle = H_{SG}|\phi_n \rangle$). If the symmetry is on-site, these operators $H_{SG}$ are sums of single-site terms, meaning that $H_{SG}$ satisfies the conditions on $H_1$ in the no-go theorem. Scarred Hamiltonians provided by this construction using on-site symmetries cannot generate entanglement dynamics on superpositions of scars, ruling out entanglement dynamics in a vast set of models. Even when an on-site symmetry is $q$-deformed \cite{odea2020tunnels} so that some of the symmetry generators are no longer on-site, the analogue of the Cartan subalgebra used to split the scars in energy is still spanned by sums of single-site operators.

\mysection{Further connections to the literature}
Our no-go theorem works for wide classes of models by inspection, but we note that a proof of a conjecture in~\cite{moudgalya2023exhaustive} will further strengthen our no-go theorem for several families of scars.

Consider a model $H$ with scar states $\ket{\phi_n} \propto (S^+)^n |\downarrow \downarrow... \downarrow\rangle$ such that $H\ket{\phi_n} = (E_0 + n \Delta)\ket{\phi_n}$ for some $E_0$ and $\Delta$ (i.e. $H$ splits the scars linearly in energy). For such a model, $H_1 = \Delta \sum_j S^z_j$ satisfies the conditions of the no-go theorem and prevents scar entanglement dynamics.

Remarkably, conjecture III.2 of~\cite{moudgalya2023exhaustive} holds that all local Hamiltonians with these scars as eigenstates indeed split the scars linearly in energy. If the conjecture holds, this means our no-go theorem immediately forbids scar entanglement dynamics in \textit{all} local Hamiltonians with the scars $\ket{\phi_n} \propto (S^+)^n |\downarrow \downarrow... \downarrow\rangle$. This conjecture is motivated by results on commutant algebras and is expected to hold in other models with scars from symmetries (see table I in~\cite{moudgalya2023exhaustive}).

We also want to highlight the work~\cite{rozon2024broken} on ``broken unitaries," which presents a related picture of replacing the time-evolution operator $e^{-i H t}$ with a simpler one on the scar subspace. ~\cite{rozon2024broken} aims to replace $e^{-i H t}$ with a product of unitaries built of terms or sums of terms of $H$ (e.g. the AKLT time evolution reduces to sequential time evolution generated by the odd and even terms $e^{-i H_{\rm AKLT} t} = e^{-i H_{\rm odd} t} e^{-i H_{\rm even} t}$ on the scar states). In the current work, we make more radical replacements where e.g. in the case of AKLT, the action of $e^{-i H t}$ on the scar states is reproduced by $e^{-i S^z_{tot} t}$ in the main text and $e^{-2 i H_{\rm even} t}$ in Appendix~\ref{app:AKLT}; $S^z_{tot}$ is not a term of $H_{\rm AKLT}.$ 

\mysection{Oscillations in deformed models}
Our second result concerns a method to evade our no-go theorem: entanglement dynamics may be induced via unitary deformations. Consider a Hamiltonian $H$ which hosts a set of scars $\ket{\phi_n}$; suppose that there exists an $H_1$ which by the no-go theorem forbids entanglement dynamics. Deform $H$ by a unitary $U$; $H \to \tilde{H} = U H U^\dagger$ and  $\ket{\phi_n} \to \ket{\tilde{\phi}_n} = U \ket{\phi_n}$. However, $H_1 \to U H_{1} U^\dagger$ will generically no longer be a sum of single-site operators, and there will generically \textit{not be any} sum of single site operators that reproduces the spectrum of $\tilde{H}$ on $\ket{\tilde{\phi}_n}$. As a result, the conditions of the no-go theorem will no longer be satisfied. 

There is another way to think about this deformation. The time evolution of $\ket{\tilde{\psi}} = \sum_n c_n \ket{\tilde{\phi}_n}$ under $\tilde{H}$ and the time evolution of $\ket{{\psi}} = \sum_n c_n \ket{{\phi}_n}$ under $H$ are related by
\begin{equation}\label{eq:tilderelation}
    \ket{\tilde{\psi}(t)} \equiv e^{-i \tilde{H} t} \ket{\tilde{\psi}} = Ue^{-i {H} t} \ket{{\psi}}
\end{equation} 
Under the assumptions on $H$, the state $e^{-i H t} \ket{\psi}$ will have static entanglement. On the other hand, $U e^{-i H t} \ket{\psi}$ can have time-varying entanglement: $U$ will generically change the entanglement of $e^{-i H t} \ket{\psi}$ at different times $t$ by different amounts. 

Furthermore, suppose $H$'s scars form an equally spaced tower and so $e^{-i H t} \ket{\psi}$ has some period $T$. By Eq.~\ref{eq:tilderelation}, $\ket{\tilde{\psi}(t)}$ will also be invariant under $t \to t+T$. As a consequence, the entanglement dynamics of $\ket{\tilde{\psi}(t)}$ are also invariant under $t \to t+T$ and hence have a period $T/n$ for some integer $n$. Since the entanglement dynamics are nontrivial and periodic, they will show oscillations and non-monotonicity in time.

As an example, we consider the scarred model introduced in \cite{markmotrunich2020}:
\begin{equation}\label{eq:DMI_model}
\begin{split}
    H_{\rm DMI} &= \sum_{i}^L \bigg[ J_1 \vec{\sigma}_i \cdot \vec{\sigma}_{i+1} + J_2 \vec{\sigma}_i \cdot \vec{\sigma}_{i+2} + h \sigma_i^z \\ &+ D \hat{z} \cdot (\vec{\sigma}_i \times \vec{\sigma}_{i+1})  \bigg]
\end{split}
\end{equation}
with scars corresponding to the $k=0$ magnon states $\ket{\phi_n} \propto (S^+)^n \ket{\downarrow ... \downarrow}$. We call this model ``DMI" because of its Dzyaloshinkskii-Moriya interaction $\hat{z} \cdot (\vec{\sigma}_i \times \vec{\sigma}_{i+1})$. The no-go theorem trivially applies with the choice of $H_1 = h \sum_i \sigma^z_i$, and so superpositions of the scars can't show entanglement dynamics under $H_{\rm DMI}$.

The state $\ket{+}^{\otimes L} = \otimes_{i=1}^L \frac{1}{\sqrt{2}} (\ket{\uparrow}_j + \ket{\downarrow}_j)$ is a superposition of all the scars and evolves to $\otimes_{j=1}^L \frac{1}{\sqrt{2}} (\ket{\uparrow}_j + e^{-2i h t}\ket{\downarrow}_j)$ (up to a global phase) under the dynamics. The entanglement is static in time (the state is unentangled at all times), and the state returns to itself after a period of $T = \frac{\pi}{h}$. 

We choose the deformation unitary $U = \exp \left(i \frac{\pi}{4} \sum_{i=1}^L \sigma_i^x \sigma_{i+1}^x\right)$. For $e^{-i h t} \neq 1,-1$, $U$ induces entanglement when it acts on $\otimes_{j=1}^L \frac{1}{\sqrt{2}} (\ket{\uparrow}_j + e^{-2i h t}\ket{\downarrow}_j)$. We chose $U$ to preserve the easily preparable scar superposition $\ket{+}^{\otimes L}$, so this state will remain a superposition of the scars in the deformed model and hence a good initial state. The half-chain entanglement entropy is straightforward to calculate and oscillates between $0$ and $\text{max}(S_{L/2}(t)) = 2\log(2)$ with a period of $\frac{\pi}{2h}$.

The deformations $U$ come with a cost. For $U$'s that can be represented as quantum circuits with local gates, the range of terms in $\tilde{H}$ grows linearly in the depth of $U$, and so only short-depth $U$ will give physical Hamiltonians $\tilde{H}$. If $U$ is given by $e^{-i \epsilon H'}$ for some local Hamiltonian $H'$ built of non-commuting terms, then even for small $\epsilon$, the terms in $\tilde{H}$ will be quasilocal with exponentially decaying tails. $U$ will also deform the scar states; a finite-depth $U$ will only change their entanglement by an $O(1)$ amount, so if the scars are distinguished by log-law entanglement~\cite{moudgalya2018entanglement,turnerRydbergscars_2018, iadecola2019xy,iadecola2020kinetic, alhambra2020revivals}, then the scars will still have logarithmic entanglement after deforming by $U$. Note that if $U$ is not chosen to preserve the initial state, then the new initial state will in general have a different baseline entanglement entropy.
\begin{figure}[h!]
    \centering
    \includegraphics[width=\linewidth]{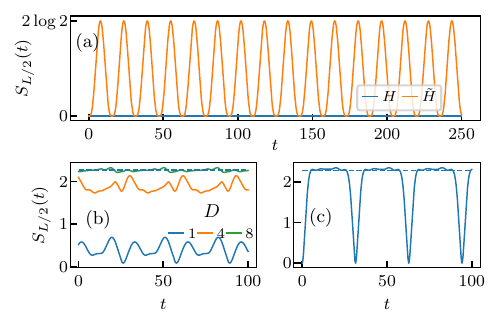}
    \caption{Entanglement dynamics in deformed DMI models $\tilde{H} = U H_{\rm DMI} U^\dagger$ for several choices of $U$. The initial state is  $U\ket{+}^{\otimes L}$, corresponding to a superposition of $\tilde{H}$'s scars. For $H_{\rm DMI}$ in Eq.~\ref{eq:DMI_model}, we use $L = 8, J_1 = 1, J_2 = -0.6, D = 0.4$ and $h = 0.1$. (a) For $U = \exp \left( i \frac{\pi}{4}\sum_i^L \sigma_i^x \sigma_{i+1}^x \right)$, the half-chain entanglement shows clear oscillations. The entanglement of the unrotated system $H=H_{\rm DMI}$ is frozen. (b) $U$ is a random circuit of depth $D$. High depth circuits increase the total entanglement towards that of a Haar state, but need not increase the magnitude of the entanglement oscillations. (c) Oscillations after rotation by unitary $U$ in eq. \ref{eq:productHaar}.  }

    \label{fig:DMI_scars}
\end{figure}

In Fig. \ref{fig:DMI_scars}(b), we show entanglement oscillations of an $L = 8$ DMI model after a transformation by depth $D = 1, 4, 8$ local random unitary circuits. As the circuit depth increases, the minimum and average values of the entanglement increase, but the extent $\max_t S_{L/2}(t) - \min_t S_{L/2}(t)$ of the oscillations is not monotonically growing in circuit depth and in fact becomes quite small at large depth. This may be intuitively understood by the fact that as circuit depth increases, $U$ becomes closer to a Haar-random unitary and $U e^{-i H t} |\psi\rangle$ will look like a Haar-random state. The entanglement entropy of Haar-random states shows only very weak fluctuations about the average value, so despite always having large entanglement, the periodic fluctuations in time will be small. Indeed, when the circuit depth is equal to $L$, we observe that the entanglement periodically fluctuates about the value of the entanglement entropy for a Haar random state.\footnote{Strictly speaking, for a local Haar circuit of depth equal to $L$, the state is not yet Haar distributed. We expect that fluctuations in entanglement will in fact keep shrinking with circuit depth up to a $\sim e^{cL}$ depth~\cite{cotler2022fluctuations}. However, the fluctuations are already strikingly small.} It is reasonable to expect that only fine-tuned unitaries $U$ can lead to large-magnitude oscillations in the entanglement entropy.

A particularly striking example of this may be seen through the following unitary:
\begin{align}\label{eq:productHaar}
    U = 
\left( \begin{array}{c|c}
\ket{+}^{\otimes_L}\bra{+}^{\otimes_L} & 0 \\
\hline
0 & \mathbf{U}_{\rm Haar}
\end{array} \right).
\end{align}
This unitary preserves the initial condition but is Haar random on the remainder of the Hilbert space. Oscillations due to this unitary are shown in Fig. \ref{fig:DMI_scars}(c). Over the course of the time evolution, the state nearly reaches the entanglement of a Haar random state. When the revival to the initial state occurs, the entanglement drops all the way down to 0. As a result, these dynamics essentially oscillate between a product state and a Haar random state. We share this example to show the power of choosing a unitary $U$ that preserves the simple product initial state of $H$; note that this example is not physically realizable, as the resulting deformed Hamiltonian $\tilde{H}$ will not be a sum of local operators.

\mysection{Summary and outlook} We have presented a framework for understanding entanglement dynamics in scarred models, giving both a no-go theorem and a method to evade the no-go theorem.  The no-go theorem is easy to apply; if there's a simple Hamiltonian (sum of single-site operators) $H_1$ that reproduces the scars' energies in some scarred Hamiltonian $H$, then $H$ can't generate scar entanglement dynamics. This $H_1$ is sometimes (e.g. DMI) but not always (e.g. AKLT) a term in the Hamiltonian $H$. The no-go theorem applies to a large number of models with exact scars, freezing the entanglement of superpositions of scars.

On the other hand, our method for evading the no-go theorem means that scarred Hamiltonians with scar entanglement oscillations are at least as numerous as those without; deforming a model without scar entanglement dynamics will generically give one with such dynamics. However, there is a corresponding cost in that the range of the terms in the deformed model will generically be larger than that of the undeformed model, and the terms themselves may possibly be less natural. This points to a two-step construction: deform a model without scar entanglement dynamics and then truncate the range of the terms. Such a model will generically lose perfect revivals but may maintain non-monotonicity of the entanglement (the entanglement may show oscillations on top of a linear growth). This would be in line with the phenomenology seen in the PXP model in the original Rydberg atom experiment; indeed, PXP has been understood~\cite{choi2019perfect} as a truncation of an unknown but approximately constructed quasilocal Hamiltonian with exact scars.

\mysection{Acknowledgements} Nicholas O'Dea thanks Wen Wei Ho for a discussion of entanglement oscillations that formed the seed for this work, and he thanks Sanjay Moudgalya for pointing out interesting connections to related literature. We thank Vedika Khemani for encouraging us to publish this work. 

Nicholas O'Dea and Adithya Sriram acknowledge funding through Vedika Khemani's Packard Fellowship in Science and Engineering and her award from the US Department of Energy, Office of Science, Basic Energy Sciences, under Early Career Award Nos. DE-SC0021111.  Adithya Sriram also acknowledges support from the National Science Foundation Graduate Research Fellowship. 

\bibliography{scarbib}

\appendix

\section{More details about AKLT}\label{app:AKLT}
In the main text, we showed that superpositions of the AKLT scar states $(Q^\dagger)^n\ket{g}$ do not show entanglement dynamics under the AKLT Hamiltonian (or any other Hamiltonian that splits these scars equally in energy): the no-go theorem is satisfied by $\hone = \sum_i S^z_i$. In this appendix, we discuss a larger set of scar states in the AKLT model to illustrate both the breakdown of the no-go theorem and simple extensions of the no-go theorem that still partially restrict the entanglement dynamics. 

\begin{figure}[t]
    \centering
    \includegraphics[width=\linewidth]{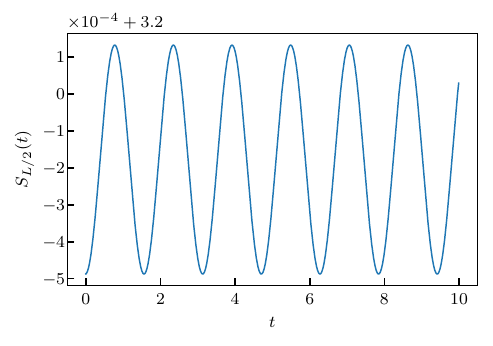}
    \caption{Half-chain entanglement entropy in the $L=10$ spin-1 AKLT model starting from a random superposition of scar states $(S^-)^m (Q^\dagger)^n\ket{g}$. Note that the magnitude of the oscillations is orders of magnitude smaller than the mean. }
    \label{fig:aklt_oscillations}
\end{figure}

By virtue of the $SO(3)$ symmetry of the AKLT model, all rotations of the scars $(Q^\dagger)^n\ket{g}$ are also eigenstates. The scar states $(Q^\dagger)^n\ket{g}$ have total spin $s = 2n$ and $\sum_i S^z_i = 2n$, so the states $\ket{\phi_{m,n}} = (S^-)^m (Q^\dagger)^n\ket{g}$ for $0 \leq m \leq 4n$ are all eigenstates of the AKLT model with energy $2n$.

This broader class of states can evade the no-go theorem, as $\hone \propto \sum_i S^z_i$ will fail to distinguish states with fixed $\sum_i S^z_i$ (i.e. states with the same value of $2n-m$), as shown in Fig. \ref{fig:aklt_oscillations}. However, there are still useful constraints that we can make in the spirit of the no-go theorem. First, decompose $H_{\rm AKLT}$ into its odd and even bonds:
\begin{equation}
\begin{split}
    H_{\rm even} &= \sum_{i=1}^{L/2} P^{(2)}_{2i,2i+1} \\
    H_{\rm odd} &= \sum_{i=1}^{L/2} P^{(2)}_{2i-1,2i}
\end{split}
\end{equation}
From appendix C of \cite{odea2020tunnels}, the alternating AKLT model annihilates the scars $(Q^\dagger)^n\ket{g}$; $(H_{\rm even} - H_0) (Q^\dagger)^n\ket{g} = 0$. Accordingly, $2 H_{\rm even} (Q^\dagger)^n\ket{g} = 2H_0 (Q^\dagger)^n\ket{g} = H_{\rm AKLT} (Q^\dagger)^n\ket{g}$. Since $H_{\rm even}$ and $H_{\rm odd}$ share the same $SO(3)$ symmetry as $H_{\rm AKLT}$, this equality holds for all the states related by $SO(3)$ symmetry: $2 H_{\rm even} (S^-)^m(Q^\dagger)^n\ket{g} = 2H_0 (S^-)^m(Q^\dagger)^n\ket{g} = H_{\rm AKLT} (S^-)^m(Q^\dagger)^n\ket{g}$.

In particular, we see that $2H_{\rm even}$ (likewise $2H_{\rm odd}$) satisfies the conditions of the generalization of the no-go theorem in the main text: it is a sum of commuting two-body operators, and so the entanglement entropy of a contiguous region will have oscillations bounded by at most $\max_t S(t) - \min_t S(t) \leq 2 \log(2)$.

We can say more. The entanglement between any even-length contiguous region of sites and the rest of the system will be unchanging. For example, suppose the region $A$ of interest was a contiguous region of size $r$ between sites $2j$ and $2(j+r)-1$. Note that $U(t) = e^{-i H_{\rm AKLT} t}$ reduces to $e^{-2 i H_{\rm even} t}$ on the space spanned by $(S^-)^m(Q^\dagger)^n\ket{g}$, and $e^{-2 i H_{\rm even} t}$ factorizes between $A$ and its complement as no operators in $H_{\rm even}$ straddle the boundaries between them. Thus, any state that is a superposition of the scars $(S^-)^m(Q^\dagger)^n\ket{g}$ will have static entanglement between any such region $A$ and its complement. Using $H_{\rm odd}$, a similar statement holds for regions $A$ between $2j$ and $2(j+r)$, so all even-length contiguous regions have frozen entanglement with their complements.

Finally, we note that even when considering an odd-length region of sites to avoid the freezing described above, the amplitude of the entanglement oscillations appears to be strikingly small compared to the mean entanglement (Fig.~\ref{fig:aklt_oscillations}). The AKLT model thus furnishes an example where despite evading no-go theorems and having entanglement oscillations, the magnitude of the oscillations are quite small.

\end{document}